\documentstyle[preprint,eqsecnum,aps,psfig]{revtex}
\tightenlines
\def\beq{\begin{equation}}
\def\eeq{\end{equation}}
\def\bea{\begin{eqnarray}}
\def\eea{\end{eqnarray}}
\def\nnu{\nonumber}
\def\tst{\textstyle}

\def\fno#1{Fig.~\ref{fig#1}}
\def\cno#1{\cite{#1}}
\def\rno#1{Ref.~\cite{#1}}

\def\al{\alpha}

\def\eps{\epsilon}
\def\tta{\theta}
\def\kap{\kappa}
\def\lam{\lambda}
\def\sig{\sigma}
\def\om{\omega}

\def\Gam{\Gamma}
\def\Dta{\Delta}
\def\Lam{\Lambda}
\def\Sig{\Sigma}
\def\Om{\Omega}
\def\Tta{\Theta}

\def\ptl{\partial}

\def\hf{{1\over2}}
\def\tshf{\tst\hf}
\def\quar{{1\over 4}}

\def\sumpr{\mathop{{\sum}'}}

\def\lp{\left(}
\def\rp{\right)}

\def\ham{{\cal H}}
\def\ket#1{|#1\rangle}

\def\tran#1#2{\langle#1|#2\rangle}

\def\mel#1#2#3{\langle#1|#2|#3\rangle}

\def\bH{{\bf H}}
\def\bJ{{\bf J}}

\def\xhat{\bf{\hat x}}
\def\zhat{\bf{\hat z}}

\def\Fe8{Fe$_8$}
\def\hsc{{\ham_{\rm sc}}}
\def\tbt{{3\over 2}}
\def\baJ{\bar J}

\def\mmi{m_r}
\def\comp{{\rm c.c.}}
\def\baj{\bar J}
\def\mo#1{\mu^2_{#1}}
\def\intrt{\int_{\mu_1}^{\mu_0}} 
\def\intcen{\int_0^{\mu_1}} 
\def\rta{\sqrt{1-h_x^2}}

\def\rtl{\sqrt\lam}
\def\rtlb{\sqrt{1 - \lam}}
\def\rtd{\sqrt{(1-h_x^2)(1-\lam)}}
\def\bk{\bar\kap}
\def\bq{{\bar q}_1}
\def\bc{\bar\chi}
\def\bto{{\bar t}_1}
\def\btt{{\bar t}_2}

\begin{document}
\draft

\title{Quenched Spin Tunneling and Diabolical Points
in Magnetic Molecules: I. Symmetric Configurations}

\author{Anupam Garg$^*$}
\address{Department of Physics and Astronomy, Northwestern University,
Evanston, Illinois 60208}

\date{\today}

\maketitle

\begin{abstract}
The perfect quenching of spin tunneling that has previously been
discussed in terms of interfering instantons, and has recently been
observed in the magnetic molecule \Fe8, is treated using a discrete
phase integral (or Wentzel-Kramers-Brillouin) method. The simplest
model Hamiltonian for the phenomenon leads to a Schr\"odinger
equation that is a five-term recursion relation. This recursion
relation is reflection-symmetric when the magnetic field applied to
the molecule is along the hard magnetic axis.  A completely general
Herring formula for the tunnel splittings for all reflection-symmetric
five-term recursion relations is obtained. Using connection formulas
for a new type of turning point that may be described as lying
``under the barrier", and which underlies the oscillations in the
splitting as a function of magnetic field, this Herring formula is
transformed into two other formulas that express the splittings in
terms of a small number of action and action-like integrals. These
latter formulas appear to be generally valid, even for problems where
the recursion contains more than five terms. The results for the model
Hamiltonian are compared with experiment, numerics, previous instanton
based approaches, and the limiting case of no
magnetic field.
\end{abstract}

\widetext

\section{Introduction}
\label{Intro}
In several previous papers \cite{opus}, we have discussed the
tunneling of a spin described by the model Hamiltonian
\beq
\ham  = -k_2 J_z^2 + (k_1 - k_2) J_x^2 - g\mu_B \bJ\cdot\bH, \label{ham} 
\eeq
where $\bJ$ is dimensionless spin operator with components
$J_x$, $J_y$, and $J_z$, $\bH$ is an external magnetic field,
and $k_1 > k_2 >0$. Since most of the earlier work has been for the
special case where $\bH\|\xhat$, let us first limit ourselves to that.
Viewed as a function of a classical angular momentum vector $\bJ$
of fixed length $J$, this Hamiltonian has two degenerate minima
for $H_x < H_c = 2k_1 J/g\mu_B$. On general grounds we expect
quantum mechanical tunneling to lift the degeneracy, and split
the energies. The surprise is that the ground state
tunnel splitting, $\Dta_0$,
is an oscillatory function of $H_x$, vanishing exactly whenever
\beq
{H_x \over H_c} = {\sqrt{1-\lam} \over J}
                  \left[J - n - \hf \right], \label{quench}
\eeq
where $\lam = k_2/k_1$, and $n = 0, 1, \ldots, 2J-1$.

The quenching of $\Dta_0$ was first studied purely as a theoretical
curiosity, and explained in terms of instantons \cite{opus}(a). 
Since then, this effect has been observed \cite{ws} in the
magnetic molecule
[(tacn)$_6$Fe$_8$O$_2$(OH)$_{12}$]$^{8+}$ (or just \Fe8 for short),
which is approximately described by the Hamiltonian (\ref{ham}),
with $J=10$, and $k_1 \approx 0.33$ K, and $k_2 \approx 0.22$ K
\cite{cs,alb,rc}. Motivated by a desire to use only elementary
methods of analysis, the problem was restudied \cite{opus}(d)
using the discrete phase integral (DPI) method (also known as
the discrete Wentzel-Kramers-Brillouin method). An approximate version
of this method was developed and applied to \Fe8 independently by
Villain and Fort \cite{vf}. The results of these later studies
confirm Eq.~(\ref{quench}) for the ground pair quenching points,
and also find additional quenching points, as we discuss next.

To help grasp the full richness of the 
spectrum of the Hamiltonian (\ref{ham}), we show in \fno1
the results of numerical calculation of the energies as a function of
$H_x$, for $J=3$, for three different values of $H_z$. In all three cases,
$H_y = 0$. In part (a), $H_z = 0$, and we have the symmetric situation
mentioned above. Note that (i) the lowest two energy level curves cross
six times (including negative values of $H_x$), and (ii) the crossing
points are perfectly periodically spaced, in complete accord with
Eq.~(\ref{quench}). Wernsdorfer and Sessoli \cite{ws}
have shown the existence of analogous crossings for \Fe8. To quantitatively
account for the observed period, one must include higher order anisotropies
in the Hamiltonian. This does not change the basic physics. In addition,
\fno1(a) also shows a number of crossings of higher energy levels, of which
the analog in \Fe8 has not been seen yet.

In \fno1(b), $H_z$ has a specific non-zero value. The problem
is no longer symmetric, and one of the classical minima is lower than the other.
Correspondingly, we see that the lowest quantum mechanical state is always
non degenerate. However, ignoring tunneling for the moment, the first excited
state in the deeper well can have an energy equal to that of the lowest
state in the
shallower well if $H_z$ is correctly chosen. And indeed, we see from the figure
that the second and third energy levels do cross a number of times. 
These crossings, when $\bH$ has an easy component, were not anticipated
in \rno{opus}(a), and were discovered by
Wernsdorfer and Sessoli experimentally. As seen in the experiments,
the crossings in \fno1(b) are shifted by half a period from
those in \fno1(a). Note that as in part (a), \fno1(b)
displays crossings between yet higher energy levels
(the fourth and fifth, e.g.),
which have also not been seen experimentally yet.

This pattern continues as $H_z$ is increased still further [\fno1(c)]. Now the
lowest two levels in the deeper well are nondegenerate, and the lowest crossings
are between levels 3 and 4. Comapred to \fno1(b), these crossings
are shifted by yet another half-period, just as seen experimentally.
Again, there are crossings between higher pairs
of levels, and again only those between levels 3 and 4 have been seen in
\Fe8.

It is clearly interesting to understand the structure in the energy spectrum
analytically, and numerical diagonalization alone cannot provide this.
When $J$ is of order 10, as it is for \Fe8, a
semiclassical analysis is natural, and it is profitable to think
of the energy differences amongst low lying levels in terms of tunneling.
Such an analysis was done in \rno{opus}(a--d). In this paper, we shall
elaborate on our earlier DPI analysis \cno{opus}(d), and provide several
results that are more
generally applicable to Hamiltonians other than Eq.~(\ref{ham}).
We shall limit ourselves, however, to problems which are analogous
to symmetric double-well
potentials in the continuum WKB case. In the context of Eq.~(\ref{ham}), this
means that $\bH\|\xhat$. The cases where $H_y$ or $H_z$ are also nonzero
correspond to an asymmetric potential, and will be considered in a second paper.

Before describing the results of our analysis, however, let us digress to make
two points. The first is the issue of degeneracy and its connection with
symmetry in light of the von Neumann-Wigner theorem.  When $\bH\|\xhat$, or 
$\bH\|\zhat$, $\ham$ is invariant under a $180^{\circ}$ rotation about $\xhat$
or $\zhat$, so energies of levels that
are odd and even under this operation can intersect. The quenchings for
$\bH\|\xhat$ [\fno1(a)] can be understood as instances of this
phenomenon \cno{opus}(b).
When $\bH$ has both $\xhat$ and $\zhat$ components, however, $\ham$ has no
symmetry, and the level crossings [\fno1(b), (c)] are nontrivial instances of
conical intersections or diabolical points \cite{hlh,bw}. 
Viewed in the larger $H_x$--$H_z$ plane, or the full three-dimensional
space of magnetic fields $\bH$, however, {\it all} points of
degeneracy are diabolical.

The second point is that there are several other special features in the
spectrum, which are evident from numerical analysis for several different $J$,
and can also be seen in \fno1.
First, the successive half-period shifts in the crossing fields as we go
from (a) to (b) to (c) in \fno1 mean that the diabolical points 
form part of a centered rectangular lattice in the $H_x$--$H_z$ plane.
The length of the rectangular unit cell along $H_x$ can be read off
Eq.~(\ref{quench}), while that along $H_z$ is given by \cite{vf,aglt}
\beq
\Dta H_z = {\lam^{1/2} \over J}H_c, \label{Hzperiod}
\eeq
where $\lam = k_2/k_1$. Second, at a diabolical point, we often find
{\it simultaneous} degeneracy of more than one pair of levels to very high
accuracy if not exactly. All these facts are captured by the DPI analysis.
In fact, in the case $\bH\|\xhat$,
all the available evidence to date --- exact diagonalization for
small $J$, perturbation theory in $\lam \equiv k_2/k_1$, numerics --- indicates
that the simultaneity of the degeneracy of many pairs of levels, as well as
the values of the degeneracy fields, are {\it exactly} given by the leading
semiclassical analysis, i.e., Eq.~(\ref{quench}) \cite{opus}(e).
These facts point to the existence of a higher dynamical symmetry, but that
is not yet established. Further, when higher order anisotropy terms
are included in the Hamiltonian to obtain quantitative agreement with
experimentally observed period \cite{ws}, the numerical evidence indicates
that although the simultaneous degeneracy of several pairs of levels and
the perfect lattice of diabolical points are no longer exact properties,
they continue to hold to rather good approximation \cite{ws2}.

The plan of the paper is as follows. In Sec.~\ref{sum} we outline the DPI
approach. The basic idea is that in the $J_z$ basis, Schr\"odinger's
equation for Eq.~(\ref{ham}) has the form of a recursion relation or difference
equation, as opposed to a differential equation for a massive particle in a
one dimensional potential $V(x)$. This difference equation can be solved in
close analogy with the continuum WKB approximation. We will see that
compared to previous DPI studies \cite{dm,sg,pab,vs} new types of turning
points arise in the study of Eq.~(\ref{ham}), because the recursion
relation has five terms as opposed to three in the earlier studies. These
turning points have no continuum analogue. Our present
discussion will rely on physical arguments and correspondence with the
continuum case.  A more formal discussion is given in \rno{jmp2}.

In Sec.~\ref{herr} we develop an analogue of Herring's formula \cite{ch,gp} for
problems leading to five term recursion relations. In the continuum case,
for a symmetric double well potential [$V(-x) = V(x)$], this formula
expresses the splitting for the $n$th pair of levels in terms of the
$x=0$ values of the wavefunction
$\psi_n(x)$ and its derivative $\psi'_n(x)$ for the $n$th state localized in one
of the wells. In Sec.~\ref{guts} we will use the DPI method to find the
analogous discrete wavefunction near the center of the potential, and use our
Herring formula to obtain a completely general formula [See Eq.~(\ref{Dgen3})]
that applies to any eigenvalue problem in the form of a recursion relation. This
latter formula is written in terms of an action integral that runs between turning
points, in close analogy with the continuum case. This formula is inconvenient
for practical applications, however, and so in subsection E of Sec.~\ref{guts},
we will transform it into another result [Eq.~(\ref{Dgen4})] that only requires
the evaluation of a small number of much simpler integrals. The second
formula is
also completely general, and has the advantage of making the $J\to\infty$
asymptotic structure of the splittings transparent. In Sec.~\ref{specific},
we will apply this lattermost formula to the \Fe8 problem, and obtain the
splitting for all pairs of levels.   We will discuss the quenching points
and several other aspects of our results, including comparison with numerics,
work by other authors \cite{klzp,lkpp}, and the features that appear to be
exact.

\section{Summary of the DPI Method}
\label{sum}
The starting point is to write Schr\"odinger's equation in the $J_z$ basis.
Suppose $\ket{\psi}$ is an eigenstate of $\ham$ with energy $E$. Then
with
$J_z \ket m = m\ket m$, $\tran{m}{\psi} = C_m$,
$\mel{m}{\ham}{m} = w_m$, and $\mel{m}{\ham}{m'} = t_{m,m'}$ ($m \ne m'$),
we have
\beq
\mathop{{\sum}'}_{n = m-2}^{m+2} t_{m,n} C_n + w_m C_m = E C_m, \label{Seq}
\eeq
where the prime on the sum indicates that the term $n=m$ is to
be omitted. The diagonal terms ($w_m$) arise
from the $J_z^2$ part of $\ham$, the
$t_{m,m \pm 1}$ terms from the $J_xH_x$ part, and the
$t_{m, m\pm 2}$ terms from the $J_x^2$ part.

We can think of Eq.~(\ref{Seq}) as a tight binding model for an
electron in a one-dimensional lattice with sites labelled by $m$,
and slowly varying on-site
energies ($w_m$), nearest-neighbor ($t_{m,m\pm 1}$), and
next-nearest-neighbor ($t_{m,m\pm 2}$) hopping terms. Since we can
think of dynamics in this model in terms of wavepackets, it is clear
that there is a generalization of the usual continuum quasiclassical
or phase integral method to the lattice case. This is the DPI method.

Previous work with the DPI method \cite{dm,sg,pab,vs} has been limited to
the case where the recursion relation has only three terms. New features
arise when five or more terms are considered. In particular, we encounter
nonclassical turning points, i.e., turning points at $m$ values other
than those at the limits of the classically allowed motion. It is these
turning points that give rise to oscillatory tunnel splittings, so that
this effect is absent in systems described by three-term recursion
relations.

The general formalism of this method \cite{pab} and the extension to
five terms is discussed at length elsewhere \cite{jmp2},
so here we will only
give a brief summary. The fundamental requirement for a quasiclassical
approach to work is that $w_m$
and $t_{m,m\pm \al}$ ($\al = 1,2$) vary slowly enough with $m$ that
we can find smooth continuum approximants $w(m)$ and $t_{\al}(m)$,
such that whenever $m$ is an eigenvalue of $J_z$, we have
\bea
w(m) &=& w_m, \label{wcont} \\
t_{\al}(m) &=& (t_{m,m+\al} + t_{m,m-\al})/2, \quad \al=1,2.
   \label{tcont}
\eea
We further demand that
\beq
{dw \over dm} = O\left( {w(m) \over J } \rp, \quad
{dt_\al \over dm} = O\left( {t_{\al}(m) \over J } \rp,
\label{slow}
\eeq
with $m/J$ being treated as quantity of order 1. We will see that
for Eq.~(\ref{ham}), these conditions will hold in the semiclassical
limit $J \gg 1$.

Given these conditions,
the basic approximation, which readers will
recognize from the continuum case, is to write the
wavefunction as a linear combination of the quasiclassical forms
\beq
C_m \sim {1 \over \sqrt{v(m)}}\exp\lp i\int^m q(m') dm'\rp,
    \label{Cwkb}
\eeq
where $q(m)$ and $v(m)$ obey the equations
\bea
E &=& w(m) + 2t_1(m) \cos q + 2t_2(m) \cos(2q)
     \equiv \hsc(q,m), \label{hjeq} \\
v(m) &=& \ptl \hsc/\ptl q = -2\sin q(m)
           \big(t_1(m) + 4 t_2(m) \cos q(m)\bigr).
     \label{vm}
\eea
Equations (\ref{hjeq}) and (\ref{vm}) are the lattice analogs of the
eikonal and transport equations. Equation (\ref{Cwkb}) represents
the first two terms in an expansion of $\log C_m$ in powers of
$1/J$.

As in the continuum case, the approximate DPI wavefunction is invalid
at turning points. 
These points arise whenever the velocity $v(m)$ vanishes for given
energy $E$, for then the approximation (\ref{Cwkb}) diverges.
We see from Eq.~(\ref{vm}) that $v(m)$
can vanish either because $\sin q =0$, i.e., $q=0$ or $q=\pi$, or
because $q = q_* \equiv \cos^{-1}(-t_1/4t_2)$. Substituting these values of
$q$ in the eikonal equation, we see that a turning point is obtained whenever
\beq
E = U_0(m),\ U_{\pi}(m),\ {\rm or}\ U_*(m),
   \label{Econd}
\eeq
where
\bea
U_0(m) &=& \hsc(0,m) = w(m) + 2t_1(m) + 2t_2(m), \label{U0} \\
U_{\pi}(m) &=& \hsc(\pi, m) = w(m) - 2t_1(m) + 2t_2(m), \label{Upi} \\
U_*(m) &=& \hsc(q_*,m) = w(m) - 2t_2(m) - {t_1^2(m) \over 4 t_2(m)}.
           \label{Ustar}
\eea
Note that at a turning point, both $m$ and $q$ are determined. If
we denote the values of these quantities generically
by $m_c$ and $q_c$, $m_c$
may be regarded as being fixed by Eq.~(\ref{Econd}), and $q_c$ by
the corresponding condition $q_c = 0$, $q_c = \pi$, or $q_c = q_*(m_c)$.

To understand the nature of these turning points, let us assume
that $t_1 < 0$, and $t_2 > 0$. [This is the case for the Hamiltonian
(\ref{ham}). We can always arrange for $t_1$ to be negative by means
of the gauge transformation $C_m \to (-1)^m C_m$. Thus there is
only one other case to be considered, namely, $t_1<0$, $t_2<0$.
This is discussed in \rno{jmp2}.] It then follows that
$U_{\pi} > U_0$, and that
\beq
U_0(m) - U_*(m) = {1 \over 4t_2(m)}
        \bigl(t_1(m) + 4t_2(m) \bigr)^2 \ge 0. \label{difU}
\eeq
Secondly, let us think of
$\ham(q,m)$ for fixed $m$ as an energy band curve. Then $U_{\pi}$ is
always the upper band edge, while the lower band edge is either
$U_0$ or $U_*$ according as whether $-t_1/4t_2$ is greater than or
lesser than 1. To deal with this possibility, it pays to introduce
a dual labelling scheme for all three curves $U_0$,
$U_{\pi}$, and $U_*$. We write $U_{\pi}(m) \equiv U_+(m)$,
and
\bea
U_0(m) = U_i(m),\  U_*(m) = U_-(m),\quad{\rm if}\ q_* \in (0,\pi),
                                             \label{Ui} \\
U_0(m) = U_-(m),\  U_*(m) = U_f(m),\quad{\rm if}\ q_* \not\in (0,\pi).
                                             \label{Uf} 
\eea
The subscripts $+$ and $-$ denote upper and lower band edges, while
the subscripts {\it i} and {\it f} denote {\it internal} and
{\it forbidden} respectively, since in the first case above, $U_0$
lies inside the energy band, while in the second case, $U_*$ lies outside.
As examples of these curves for a symmetric recursion relation,
we show those for \Fe8 in \fno2. A magnified view of the lower left
hand portion of this diagram is given in \fno3.

Turning points where $E=U_+$, or $E=U_-$ when $U_- = U_0$, are
analogous to those encountered in the continuum quasiclassical
method, since the energy lies at a limit of the classically allowed
range for the value of $m$ in question. Points where $E=U_-$ when
$U_- = U_*$ are physically analogous, but mathematically different
since the value of $q_c$ is neither $0$ nor $\pi$. Points where
$E = U_i$ (see the energy $E_1$ in \fno2, e.g.) are novel in that
the energy is {\it inside} the
classically allowed range for $m_c$, but the mathematical form
of the connection formulas is identical to the case $E=U_-=U_0$
since $q_c = 0$. Most interesting are the turning points with
$E = U_f$ (the point $m=-m_1$ in \fno3, for instance),
since now the energy is outside the allowed range
for $m=m_c$, and the value of $q_c$ is therefore necessarily complex.
These points lie ``under the barrier" and turn out to be the ones of
importance for understanding oscillatory tunnel splittings.

The above discussion shows that the curves $U_0$, $U_{\pi}$, and
$U_*$ collectively
play the same role as the potential energy in the continuum
quasiclassical method. We refer to them as {\it critical curves}.
We have already noted that $U_{\pi} > U_0 \ge U_*$. Let us suppose
that the case of equality in Eq.~(\ref{difU}) occurs at $m=m^*$.
Clearly $t_1(m^*)/4t_2(m^*) = -1$, which is
precisely the condition found above for the lower band edge to
change from $q=0$ to $q=q_*$. Secondly, expanding $t_1$ and $t_2$
about $m*$, we see that $U_0$ and $U_*$ have a common tangent
when they meet.

\section{Herring's formula for five-term recursion relations}
\label{herr}
The problem of computing tunnel splittings in a symmetric double
well potential in the continuum case is greatly simplified by use
of Herring's formula \cite{ch,gp}. An entirely analogous formula
can be derived in the discrete case \cite{vw,vs}(c) following the
simplified treatment of Landau and Lifshitz \cite{ll}.

We have already noted the importance of the critical curves. For low
lying energy levels, in particular, the curve $U_-$
is very much like the potential energy in the continuum case,
and it is clear that we will have an entire series of approximate
energy eigenstates with wavefunctions localized in any one of the
two wells, in the vicinity of $\pm m_0$, the minima of $U_-(m)$.
(See \fno2.) Let $C_m$ be the $n$th such wavefunction localized in
the right hand well, normalized to unit total probability,
and let it satisfy Schr\"odinger's equation
(\ref{Seq}) with an energy $E_0$ for all values of $m$ well to the
right of the left well, including in particular the region around
$m=0$. More precisely, we take $C_m$ to decay away from the right
well in {\it both} directions. Such a function could be obtained, e.g.,
as the energy eigenfunction of a modified problem in which the on-site
energy is increased by a large positive amount for all $m < m_a$,
where $-m_0 \ll m_a \ll 0$, it being understood that $m_a$ is far
away from all turning points for the energy concerned. However, this
problem need not be solved explicitly, as the exact behavior of
$C_m$ near $m=-m_0$ is never needed, and therefore need not be
examined too closely.

Given such a function, Herring shows that the true symmetric and
antisymmetric eigenfunctions, $s_j$ and $a_j$, with energies
$E_1$ and $E_2$ respectively, are given very accurately by
\beq
\begin{array}{rcl}
a_m & = & {\displaystyle{1\over \sqrt 2}} (C_m - C_{-m}), \\
\noalign{\vskip5pt}
s_m & = & {\displaystyle{1\over \sqrt 2}} (C_m + C_{-m}). \\
\end{array}
\label{ajsj}
\eeq
The product $C_m C_{-m}$ is exponentially small everywhere, these
functions are normalized to unit total probaility to exponentially
high accuracy.

The Schr\"odinger equations obeyed by $C_m$ and $a_m$ are
\bea
(w_m - E_0)C_m + \sumpr_{n = m-2}^{m+2} t_{m,n} C_n &=& 0,\label{Cm0} \\
(w_m - E_1)a_m + \sumpr_{n = m-2}^{m+2} t_{m,n} a_n &=& 0.\label{am0}
\eea
Let us now define $\mmi$ to be $1$ if $J$ is integral, and $1/2$
when $J$ is half-integral. Multiplying Eq.~(\ref{Cm0}) by $a_m$,
Eq.~(\ref{am0}) by $C_m$, and summing over $m$ from $\mmi$ to
$J$, we get
\beq
(E_1 -E_0) \sum_{m = \mmi} C_m a_m + \Sig_1 - \Sig_2 = 0,
\label{e1e0}
\eeq
where
\bea
\Sig_1 &=& \sum_{m = \mmi}^J \sumpr_{n=m-2}^{m+2}
                          a_m t_{m,n}C_n, \label{sig1} \\
\Sig_2 &=& \sum_{m = \mmi}^J \sumpr_{n=m-2}^{m+2}
                          C_m t_{m,n}a_n. \label{sig2}
\eea

To simplify Eq.~(\ref{e1e0}), we first note that by Eq.~(\ref{ajsj})
\beq
\sum_{m = \mmi} C_m a_m
      \approx {1\over\sqrt 2} \sum_{m = \mmi} C_m^2
      \approx {1\over \sqrt 2}, \label{norm}
\eeq
since the product $C_m C_{-m}$ is everywhere exponentially
small, and since $C_m^2$ is concentrated almost completely
in the right well. Secondly,
most of the terms in the sums $\Sig_1$ and $\Sig_2$ can be
seen to be identical by shifting the summation indices in various
terms suitably, and making use of the symmetry $t_{m,n} = t_{n,m}$.
For example, the difference between the terms
in $\Sig_1$ with $n = m+2$, and those in $\Sig_2$ with $n = m-2$
equals
\bea
\sum_{m = \mmi}^J \lp a_m t_{m,m+2} C_{m+2}
                - C_m t_{m,m-2} a_{m-2} \rp
   &=& \sum_{m = \mmi}^J a_m t_{m,m+2} C_{m+2}
      -\sum_{m = \mmi -2 }^{J-2} C_{m+2} t_{m+2,m} a_m \nnu \\
   &=& - a_{\mmi - 1} t_{\mmi -1, \mmi +1}C_{\mmi + 1}
       - a_{\mmi - 2} t_{\mmi -2, \mmi}C_{\mmi}, \label{difeg}
\eea
where we have made use of the obvious facts that $t_{J,J+2}$
and $t_{J-1, J+1}$ are identically zero.
The differences between the other terms in $\Sig_1$ and
$\Sig_2$ can be similarly evaluated, and reduce to a small
number of terms involving the product of an $a$ with a $C$,
which can then be written entirely in terms of $C$'s using
Eq.~(\ref{ajsj}). Finally, we can see that
$E_1 - E_0$ = $E_0 - E_2$ = $\pm\Dta/2$, and the net result is that
upto an irrelevant over all sign,
\beq
\Dta = \cases{
        2\left[ t_{0,1} C_0(C_1 - C_{-1}) + t_{0,2}C_0(C_2 - C_{-2})
                + t_{-1,1} (C_1^2 - C_{-1}^2) \right],
                                & \\integer $J$, \cr
        2\ t_{-\hf,\hf} \Bigl( C^2_{\hf} - C^2_{-\hf}\Bigr)
         + 4\ t_{-\tbt,\hf} \lp C_{\hf}C_{\tbt} - C_{-\hf}C_{-\tbt}\rp,
                                & half-integer $J$. \cr}
 \label{Herr}
\eeq

Herring gives a more careful justification of his formula by employing
the Temple-Kato error bound on energy eigenvalues \cite{gt,tk}. His
argument can be adapted word for word to the present problem, and shows
that the error in the splitting as calculated via Eq.~(\ref{Herr}) is
exponentially smaller than the splitting itself, by a factor such
as $e^{-cJ}$ where $c>0$. As $J \to\infty$, therefore, Eq.~(\ref{Herr})
is asymptotically correct.

We remind readers that Eq.~(\ref{Herr}) is not limited to the ground
state splitting.

\section{General Formula for Tunnel Splitting in terms of action
integrals}
\label{guts}
To apply Herring's formula (\ref{Herr}) to the Hamiltonian
(\ref{ham}), we will use the DPI approximation for the wavefunction.
Actually, we will take $C_m$ in Eq.~(\ref{Herr}) to be localized in
the left well. This can only change the answer by a sign, which is not
of interest to us anyway.

\subsection{DPI form near potential well minimum}

Let us first take up the problem of finding the DPI approximation
to $C_m$ in somewhat general terms. Step 1 is to find $C_m$ in the
classically allowed region near $-m_0$, the minimum of $U_-(m)$.
(See \fno3.)
We assume, as will be seen to be true for Eq.~(\ref{ham}), that in 
this region $U_- = U_0$. For energies close to $U_-(-m_0)$, and
$m$ close to $-m_0$, the
eikonal equation can only be satisfied if $q$ is close to zero.
We can therefore expand $\hsc$ in powers of $m+m_0$ and $q$:
\beq
\hsc(q,m) \approx U_-(-m_0) + {1\over 2M} q^2 +
           \hf M \om_0^2 (m+m_0)^2 + \cdots \label{hm0}
\eeq
where
\bea
M &=& - \left[2t_1(-m_0) + 8 t_2(-m_0) \right]^{-1} > 0, \label{Mass} \\
\om_0^2 &=& - \left.
            2(t_1 + 4t_2){\ptl^2 U_- \over \ptl m^2} \right|
                   _{m = -m_0}. \label{omsq}
\eea
Note that by virtue of Eq.~(\ref{slow}), and its natural extension to
second derivatives, $\om_0$ is of order $1/J$ relative to $t_1$ and
$t_2$.

The allowed eigenvalues and eigenfunctions can now be written down
very simply by noting that the eikonal equation is also the
Hamilton-Jacobi equation with
$q = \ptl\Phi/\ptl m$, $\Phi$ being the action.
Thus the problem is identical to that of a harmonic oscillator.
(Alternatively, we could arrive at the same result by approximating
the original recurrence relation by a differential equation in the
vicinity of $-m_0$.) For the $n$th state, therefore,
\beq
E_0 = U_-(-m_0) + \bigl(n+{\tshf}\bigr)\om_0, \label{e0n}
\eeq
and
\beq
C_m = \lp 2^{2n} (n!)^2 \pi\xi^2 \rp ^{-1/4}
                  e^{-x^2/2\xi^2} H_n(x/\xi),
\label{hpp}
\eeq
where $x = m+m_0$, $H_n$ is the $n$th Hermite polynomial, and
$\xi = (M\om_0)^{-1/2}$. The wavefunction is already normalized,
and the additional tails from the forbidden
region only modify the normalization by an exponentially small
amount.

It is apparent that the expansion (\ref{hm0}) is invalid unless the
point $-m_0$ is sufficiently far from the edge $m = -J$. Since the
width of the wavefunction (\ref{hpp}) is ${\sqrt n}\xi$, a necessary
condition for the validity of our procedure is
\beq
J - m_0 \gg {\sqrt n}\xi. \label{edgecond}
\eeq
If this condition does not hold, then the recursion relation must be
solved near the edge by a different method, which is tantamount to
using the Holstein-Primakoff or Bogoliubov transformation. An example
of the latter approach is given in Sec.~IV of \rno{jmp1}.

From the viewpoint of the DPI method, we have two turning
points very close to $-m_0$, one to the left, and one to the
right, since the condition $E= U_-(m)$ is then satisfied. The one to
the left has been discussed above. Let
us now consider the one to the right, and denote it by $-m_t$.
We have
\beq
-m_t + m_0 = \left[ 2n+1 \over M\om_0 \right]^{1/2}
             \sim (nJ)^{1/2}. \label{mt}
\eeq
The neglected terms in Eq.~(\ref{hm0}), on the other hand, are of
relative orders $q^4$, $(m+m_0)^3/J^3$, and $(m+m_0)q^2/J$, and
thus smaller than $n\om_0$ for $x \ll (nJ^2)^{1/3}$.
Thus, provided $n \ll J$, the solution (\ref{hpp}) holds well past
$-m_t$, and can be matched directly onto the DPI solution under the
barrier, without any need of connection
formulas at $m=-m_t$ \cite{bm}. This argument is given at greater length in
Sec.~V of \rno{jmp1}.

\subsection{DPI form in ordinary forbidden region}

Step 2 is to consider the DPI solution for $m>-m_t$. Since we
want this solution to decay as $m$ increases, we take it as
\beq
C_m = {B \over \sqrt{|v(m)|}}
          \exp\lp -\int_{-m_t}^m \kap(m') dm'\rp, \label{Cminter}
\eeq
where $\kap(m) = {\rm Im}q(m) > 0$. This solution must be matched on
to (\ref{hpp}) to determine $B$. We can continue to use the harmonic
oscillator approximation (\ref{hm0}) to $\hsc$ for this purpose, and
a simple calculation \cite{amjp}, which may in fact be traced back
to Furry \cite{wf}, leads to 
\bea
B &=& \lp{\om_0 g_n \over 2\pi}\rp^{1/2}; \label{Bgn} \\
g_n &=& {\sqrt{2\pi} \over n!} \lp n + \tshf \rp^{n +\hf}
         e^{-(n + \hf)}. \label{gndef}
\eea
The quantity $g_n$ is defined so that $g_n \to 1$ as $n\to\infty$;
$g_0 = (\pi/e)^{1/2} \approx 1.075$, $g_1 \approx 1.028$,
$g_2 \approx 1.017$, $\ldots$.

\subsection{DPI form in central region}

Step 3 is to find the wavefunction in the central region near
$m=0$. This is already done if there are no turning points
between $-m_t$ and $m=0$. For the Hamiltonian (\ref{ham}), it
turns out that we encounter another turning point where $E = U_f(m)$
(the only possibility) at an intermediate point $m=-m_1$ (see \fno3).
The solutions for $m < -m_1$ and $m > -m_1$ must therefore be related
by a connection formula. To understand this turning point, we note
that the eikonal equation (\ref{hjeq}) may be solved for
$\cos q$ as
\beq
\cos q(m) = {-t_1(m) \pm [t^2_1(m) - 4t_2(m) f(m)]^{1/2}
              \over 4t_2(m)}, \label{cosq}
\eeq
where $f(m) = w(m) - 2t_2(m) -E$. Since $\cos q = -t_1/4t_2$ at
$m=-m_1$, the discriminant in Eq.~(\ref{cosq}) must vanish, and
we conclude that as we cross $-m_1$, $\cos q$ changes from real to
complex, and $q$ changes from imaginary to complex. [Incidentally,
it may 
be verified that the condition for vanishing discriminant, i.e.,
\beq
t_1^2(m) = 4t_2(m) \bigl( w(m) - 2t_2(m) - E \bigr), \label{tp2}
\eeq
is identical to $E = U_*(m)$.] Since the recursion relation is real,
the solution $C_m$ must also be real for all $m$. A single DPI
solution for $m>-m_1$ cannot meet this demand, and so we must take
a linear combination of two DPI forms with complex conjugate $q$'s.
If we write these as
\beq
q_{1,2}(m) = i\kap(m) \pm \chi(m), \label{kapchi}
\eeq
with $\kap$ and $\chi$ both real, then we must still have $\kap > 0$
in order that $C_m$ continue decaying, and we may also take $\chi > 0$.
Let us further write the solution (\ref{Cminter}) for $m < -m_1$ as
\bea
C_m &=& {A \over 2\sqrt{|v(m)|}}
          \exp\lp -\int_{-m_1}^m \kap(m') dm'\rp, \label{Cmint2} \\
A &=& 2B \exp\lp -\int_{-m_t}^{-m_1} \kap(m') dm'\rp.  \label{BtoA}
\eea
Then, as shown in \rno{jmp2}, the DPI solution for $m > -m_1$
is given by
\beq
C_m = {\rm Re} {A \over \sqrt{s_1(m)}}
      \exp \lp i\int_{-m_1}^m q_1(m') dm' \rp,
   \label{Crite}
\eeq
where $s_1(m) = -iv\bigl(q_1(m)\bigr)$ \cite{fn1}.

\subsection{Herring's formula with DPI approximation}

The solution (\ref{Crite}) is ripe for substitution into the Herring
formula (\ref{Herr}). To do this, we first note that
\bea
\cosh\kap \cos\chi &=& -t_1/4t_2, \label{chcos} \\
\sinh\kap \sin\chi &=& \bigl(4t_2f - t_1^2\bigr)^{1/2}/4t_2. \label{shsin}
\eea
It then follows that
\beq
s_1 = 8t_2(m) \sinh\kap(m) \sin\chi(m) \sin q_1(m). \label{s1m}
\eeq
We now substitute Eqs.~(\ref{Crite})--(\ref{s1m}) into
Herring's formula,
Eq.~(\ref{Herr}). In doing this, we may neglect the variation of
quantities $t_{\al}(m)$, $q(m)$, and $v(m)$ among the sites near the
center of the lattice, since the number of sites involved is of order
$1$, and so the variation leads to higher order corrections in powers
of $1/J$. To save writing, we denote quantities evaluated at $m=0$ by
a bar: $q_1(0) \equiv \bq$, $\kap(0) \equiv \bk$, etc.
We thus get
\beq
C_m = {\rm Re} A_2 { e^{i(\Om + m \bq)}\over \sqrt{\sin \bq}},
   \label{Cmid}
\eeq
where,
\bea
\Om &=& \int_{-m_1}^0 q_1(m')dm', \label{Omdef} \\
A_2 &=& (8\btt \sinh\bk\sin\bc)^{-1/2}A. \label{AtoA2}
\eea

The cases of integer and half-integer $J$ are best tackled separately.
Doing the former first, we have
\bea
C_1 - C_{-1} &=&
    iA_2 \left[e^{i\Om} \sqrt{\sin \bq} - \comp \right], \label{jlt1}\\
C_1 + C_{-1} &=& 
     A_2 \left[e^{i\Om} \sqrt{\cos^2 \bq \over \sin \bq}
                       + \comp \right], \label{jlt2}\\ 
C_0(C_1 - C_{-1}) &=& i{A^2_2 \over 2}
     \left[\lp e^{2i\Om} - e^{-2{\rm Im}\Om}
     \sqrt{\sin \bq^* \over \sin \bq} \rp - \comp \right], \label{jlt3}\\ 
C_2 - C_{-2} &=& 
    2iA_2 \left[e^{i\Om} \cos \bq\sqrt{\sin \bq} -
                        \comp \right], \label{jlt4}\\
C_0(C_2 - C_{-2}) &=& iA^2_2 
     \left[\lp e^{2i\Om}\cos \bq - e^{-2{\rm Im}\Om}\cos \bq^*
     \sqrt{\sin \bq^* \over \sin \bq} \rp - \comp \right], \label{jlt5}\\ 
C^2_1 - C^2_{-1} &=& iA^2_2 
     \left[\cos \bq\lp e^{2i\Om} \cos \bq - e^{-2{\rm Im}\Om}
     \sqrt{\sin \bq^* \over \sin \bq} \rp - \comp \right]. \label{jlt6}
\eea
Substituting these and the formula
$\bto = -4\btt\cosh\bk\cos\bc$ into Eq.~(\ref{Herr}), we get
\bea
\Dta &=& -8A_2^2 \btt {\rm Im}\lp e^{2i\Om}\ \Tta
        -e^{-2{\rm Im}\Om} \sqrt{\sin \bq^* \over \sin \bq} 
             \ {{\rm Re}} \Tta \rp;
                    \label{Dta1} \\
\Tta &=& \cos \bq - \cosh\bk\cos\bc. \label{Tta}
\eea
But, it follows from Eq.~(\ref{kapchi}) that
\beq
\cos \bq = \cosh\bk\cos\bc - i \sinh\bk\sin\bc,
         \label{cosq1}
\eeq
so the second term in Eq.~(\ref{Dta1}) vanishes altogether, and
\bea
\Dta &=& 4 A_2^2 \btt \sinh\bk \sin\bc
           (e^{2i\Om} + e^{-2i\Om^*}) \nnu \\
     &=& \tshf A^2 (e^{2i\Om} + e^{-2i\Om^*}), \label{Dgen}
\eea
where we have used Eq.~(\ref{AtoA2}) in the last step.

For half-integer $J$, we get
\bea
C^2_{\pm 1/2} &=& {A_2^2 \over 4}
       \left[ \lp {e^{2i\Om} \over \sin \bq} e^{\pm i\bq}
          + {e^{-2{\rm Im}\Om} \over |\sin \bq|} \rp
                           +\comp \right], \label{jlt7} \\
C_{\pm 1/2} C_{\pm 3/2} &=& {A_2^2 \over 4} 
       \left[ \lp {e^{2i\Om} \over \sin \bq} e^{\pm 2i\bq}
          + {e^{-2{\rm Im}\Om} \over |\sin \bq|} e^{\mp i\bq} \rp
                           +\comp \right]. \label{jlt8} 
\eea
Thus,
\bea
C^2_{1/2} - C^2_{-1/2} &=& {i \over 2} A_2^2
                             (e^{2i\Om} - \comp), \label{jlt9} \\
C_{1/2} C_{3/2} - C_{-1/2} C_{-3/2} &=&
              iA_2^2(\cos \bq e^{2i\Om} - \comp), \label{jlt10}
\eea
and
\bea
\Dta &=& iA_2^2 \left[ (\bto + 4\btt)e^{2i\Om}
                      - \comp \right] \nnu \\
     &=& 4A_2^2 \btt \sinh\bk\sin\bc
           (e^{2i\Om} + e^{-2i\Om^*}), \label{Dgen2}
\eea
which leads, once again, to Eq.~(\ref{Dgen}).

Collecting Eqs.~(\ref{Bgn}), (\ref{BtoA}),
(\ref{Omdef}) and (\ref{Dgen}), and making use of the symmetry of
the problem, we may write the tunnel splitting
for both integer and half-integer $J$ as
\beq
\Dta = {\om_0 g_n \over 2\pi}
      \left[ \exp \lp i\int_{-m_t}^{m_t} q(m') dm' \rp
               + \ \comp \right].
     \label{Dgen3}
\eeq
Here $q(m')$ is chosen to have a positive real part $\chi$ in the
first term.
We note once again that this result applies to higher pairs
of excited states, and not just the ground pair. The essential
dependence on $n$, the excitation number, enters through the
$n$ dependence of $m_t$, the turning point.

The similarity of Eq.~(\ref{Dgen3}) to the final result in \rno{ll}
is striking \cite{fn2}, and one can ask whether one should not have
anticipated it right away. For the ground state pair, the instanton
approach \cite{opus}(a) makes it very easy to understand the
presence of two complex conjugate tunneling actions, and the fact
that they should be superposed, but does not give the prefactor.
The action integrals in the instanton approach, however, run not from
turning point to turning point but from one minimum of the energy
to the other. Further,
properly justifiying the prefactor using instantons has proven very
difficult \cite{es,bpp}. Purely as a recipe for calculations, however,
a hybrid approach, in which one adds the tunneling actions from all
equivalent instantons, and uses the DPI approach to determine the
form of the prefactor, would appear to be valid for all problems.
Thus, we strongly suspect that Eq.~(\ref{Dgen3}) is correct even when
the recursion relation has seven or more terms.

\subsection{Extraction of singular parts of action integrals}

While the formula (\ref{Dgen3}) is very general, it has the
disadvantage that the action integral runs between turning points.
The integrand is therefore close to a singularity, and
for low lying states, this gives rise to terms in the action that
depend on $\ln J$. Hence the formula does not reveal the asymptotic
behavior as a function of $J$ in a transparent way.

In this subsection, we will show that we can write the splitting 
very simply in a way that does not suffer from the above drawback.
The final result is
\beq
\Dta_n = {1\over n!} \sqrt{8\over\pi} \om_0 F^{n+\hf}
         e^{-\Gam_0}\cos\Lam_n,
      \label{Dgen4}
\eeq
where,
\bea
\Gam_0 &=& 2\int_{-m_0}^0 \kap_0(m) dm, \label{Gam0} \\
\Lam_n &=& 2 \int_{-m_1}^0 \biggl(\chi_0 + 
                         (n + \tshf)\om_0\chi'_0\biggr) dm,
                                              \label{Lamans}\\
F &=& 2M\om_0 (m-m_1)^2 \exp\lp-2\lp Q_1 +
                          \om_0\int_{-m_1}^0 \kap'_0 dm \rp\rp,
                                            \label{Fans} \\
Q_1 &=&  \int_{-m_0}^{-m_1} \lp
             {\om_0 B'_0 \over \sqrt{B_0^2 - 1}}
                + {1 \over m+m_0}\rp dm. \label{Q1def}
\eea
In Eqs.~(\ref{Gam0}--\ref{Q1def}),
the irregular turning points $\pm m_1$ may be evaluated by setting
$E = U_-(\pm m_0)$, and it should be recalled that $\pm m_0$ are the
minima of $U_-(m)$. Further,
\bea
\kap_0 &=& \kap(m,\eps = 0);\quad
   \kap'_0 = \biggl. {\ptl \kap(m,\eps) \over \ptl\eps}\biggr|_{\eps=0},
                                          \label{k0kpr0}\\
\chi_0 &=& \chi(m,\eps = 0);\quad
   \chi'_0 = \biggl. {\ptl \chi(m,\eps) \over \ptl\eps}\biggr|_{\eps=0},
                                          \label{c0cpr0}\\
B_0 &=& \cos q(m,\eps = 0);\quad
   B'_0 = \biggl. {\ptl \cos q(m,\eps) \over \ptl\eps}\biggr|_{\eps=0},
                                          \label{B0Bpr0}
\eea
with
\beq
\eps \equiv E - U_-(-m_0). \label{epsdef}
\eeq

The problem of finding the low level splittings is thus reduced to the
evaluation of a handful of integrals.  The proliferation of notation
masks the actual simplicty of these formulas. 

To derive these results, we follow a procedure similar to that used for
the continuum case in \rno{amjp}. We begin by defining
\beq
\Phi(\eps) = -i\int_{-m_t(\eps)}^0 q(m,\eps) dm, \label{Phidef}
\eeq
where the energy dependence is made explicit. The splitting for
the $n$th pair of states is then given by
\beq
\Dta_n = {\om_0 g_n \over 2\pi}(e^{-2\Phi(\eps_n)} + \comp),
             \label{Dtan}
\eeq  
with $\eps_n = (n+\tshf)\om_0$. Writing $x = m + m_0$ as in
Eq.~(\ref{hpp}), the integrand in $\Phi$ behaves as
$(x^2 - x_t^2)^{1/2}$ near the lower limit, with
$x_t = -m_t + m_0 \sim \eps^{1/2}$. Thus there is a singular
part in $\Phi$ of the form $\eps\ln\eps$, which it is our goal to
extract. To this end, we differentiate Eq.~(\ref{Phidef}) to get
\beq
\Phi'(\eps) = {d\Phi \over d\eps}
            = -i\int_{-m_t(\eps)}^0 {\ptl q \over \ptl\eps} dm.
                \label{Phipr}
\eeq
Note that the term arising from differentiating the lower limit
vanishes, nor is there any explicit contribution from the singular
behavior $q \sim (m+m_c)^{1/2}$ for $m$ near $-m_c$.

Next, let us divide $\Phi'(\eps)$ into two integrals,
$\Phi'_1$, in which the limits of integration are $-m_t$ and $-m_1$,
and $\Phi'_2$, which runs from $-m_1$ to $0$. Defining
\beq
B_{\eps}(m) = \cos\bigl( q(m,\eps)\bigr), \label{Bdef}
\eeq
we have
\beq
\Phi'_1(\eps) = \int_{-m_t(\eps)}^0
               {B'_{\eps}  \over \sqrt{B_{\eps}^2(m) - 1}} dm,
                       \label{Phip1}
\eeq
where $B'_{\eps} = \ptl B_{\eps}/\ptl\eps$.
It follows from Eq.~(\ref{hm0}) that near $m = -m_t$,
\beq
B_{\eps} \approx 1 + \lp \tshf M\om^2 x^2 - \eps \rp M + \cdots,
           \label{Bnear0}
\eeq
so the integrand in Eq.~(\ref{Phip1}) behaves as
$-1/\om_0 (x^2 - x_t^2)^{1/2}$. If we add and subtract the integral
of this expression, we obtain
\beq
\Phi'_1(\eps) = -{1 \over \om_0}
               \int_{x_t}^{x_1} {dx \over \sqrt{x^2 - x_t^2}}
              + \int_{x_t}^{x_1} \left[
               {B'_{\eps}  \over \sqrt{B_{\eps}^2(m) - 1}}
                + {1 \over \om_0\sqrt{x^2 - x_t^2}} \right] dx,
              \label{Phi1sub}
\eeq
where $x_1 = m_0 - m_1$.
The first integral can be evaluated exactly. In the second integral
we can put $\eps = 0$ both in the limits and in the integrand, since
we are not interested in terms of $O(\eps)$. Ignoring terms of this
order throughout, and making use of the relation
\beq
x_t^2 = 2\eps/M\om_0^2, \label{xtans}
\eeq
we obtain
\beq
\Phi'_1(\eps) = {1 \over 2\om_0}\left[
                 \ln{\eps \over 2M \om_0 (m_0 - m_1)^2}
                 + 2Q_1 \right], \label{Phi1ans}
\eeq
where $Q_1$ is given by Eq.~(\ref{Q1def}). Also, we can evaluate
$m_1$ at $\eps = 0$.

The remaining contribution to $\Phi'(\eps)$, $\Phi'_2(\eps)$,
can be evaluated simply by putting $\eps = 0$, since the neglected
part is $O(\eps)$. Recalling the definitions (\ref{k0kpr0}) and
(\ref{c0cpr0}), we have
\beq
\Phi'_2(\eps) \approx \int_{-m_1}^0
                      (\kap'_0 - i\chi'_0) dm.
                  \label{Phi2ans}
\eeq

We now integrate the expression for $\Phi'(\eps)$
and obtain $\Phi$. It is useful to separate the real and imaginary
parts of the answer at this stage. For the real part, we get
\beq
\Gam = 2\,{\rm Re}\Phi =
        \Gam_0 + {\eps \over \om_0}
          \left[ 2Q_1 - 1 + \ln{\eps \over 2M \om_0 (m_0 - m_1)^2}
             + 2\om_0 \int_{-m_1}^0 \kap'_0\, dm \right],
                    \label{Gamans}
\eeq
with $\Gam_0$ given by Eq.~(\ref{Gam0}), while for the imaginary part,
$\Lam_n \equiv -2\,{\rm Im}\Phi$, we get Eq.~(\ref{Lamans}).

Substituting Eqs.~(\ref{Gamans})--(\ref{Lamans}), and the definition
(\ref{gndef}) of $g_n$ in the formula (\ref{Dtan}) for $\Dta_n$, and
recalling that $\eps_n = (n+\tshf)\om_0$, we finally obtain the answer
quoted at the start, Eq.~(\ref{Dgen4}).

\section{Application to the \Fe8 problem}
\label{specific}

We now apply our general formulas to the specific problem of
\Fe8, as described by the Hamiltonian (\ref{ham}). The various
matrix elements of this Hamiltonian are given by
\bea
w_m &=& \hf(k_1 + k_2)[J(J+1) - m^2], \label{matw} \\
t_{m,m+1} &=& -\hf g \mu_B H_x [J(J+1) - m(m+1)]^{1/2},
                                      \label{matt1} \\
t_{m,m+2} &=& \quar (k_1 - k_2)
              \left[ [J(J+1) - m(m+1)][J(J+1)- (m+1)(m+2)]
               \right]^{1/2}. \label{matt2}
\eea
We must now replace these by continuous
functions $w(m)$, $t_1(m)$, and
$t_2(m)$. Since our formalism requires knowing the first two
terms in the action in an expansion in powers of $1/J$, it follows
that we need only determine the functions $w(m)$ etc. to the
same order. Furthermore, this determination need not be made in the
form of a power series, and any functional representation that
gives the first two terms correctly will be adequate. The most
convenient way to do this is to replace the combination $J(J+1)$
in the above expressions by $\baj^2$, where
\beq
\baj = J + \tshf. \label{barJ}
\eeq
The evaluation of the integrals (\ref{Phipr})--(\ref{Q1def}) is then
lengthy, but straightforward. We will present and discuss the final
results first, and give the details of the analysis later.

\subsection{Tunnel splittings for \Fe8}
\label{fe8results}
The final result for the splitting of the $n$th pair of levels is
\beq
\Dta_n = {1\over n!} \sqrt{8\over\pi} \om_0 F^{n+\hf}
         e^{-\Gam_0}\cos\Lam_n,
      \label{Dnfinal}
\eeq
where,
\bea
\om_0 &=& 2J[k_1k_2(1-h_{x0}^2)]^{1/2}, \label{om0fe8}\\
F &=& 8 J {\lam^{1/2} (1 - h_x^2)^{3/2} \over 1 - \lam - h_x^2},
        \label{Ffe8} \\
\Gam_0 &=& \baj \left[
           \ln \lp {\rta + \rtl \over \rta - \rtl} \rp
           - {h_x \over \rtlb}
              \ln \lp {\rtd + h_x \rtl \over \rtd - h_x \rtl} \rp
            \right],  \label{Sr} \\
\Lam_n &=& {{\rm max}} \left\{0,\ 
                 \pi J\lp 1 - {H_x \over \sqrt{1-\lam}H_c} \rp
              - n\pi \right\}.
   \label{Lamtot}
\eea
In Eqs.~(\ref{om0fe8}--{\ref{Lamtot}), $\lam = k_2/k_1$, and
\beq
h_x = {JH_x \over \baj H_c}, \quad h_{x0} = {H_x \over H_c}.
         \label{hxhx0}
\eeq
Recall that $H_c = 2k_1 J/g\mu_B$.

Let us now turn to the discussion of these results. The first point
concerns the fields where the $n$th tunnel splitting vanishes. Taking
account of the fact that $\Lam_n$ is necessarily positive as indicated
by Eq.~(\ref{Lamtot}), we see that this happens whenever
\cite{aglt,opus}(d,e)
\beq
{H_x \over H_c} = {\sqrt{1-\lam} \over J}
                  \left[J - \ell - \hf \right], \label{quench2}
\eeq
with $\ell = n$, $n+1$, $\ldots$, $2J - n -1$, yielding $2(J-n)$
quenching points in all for $\Dta_n$. When $n=0$, these are the results
quoted in Sec.~\ref{Intro}.

In \fno4 we compare Eq.~(\ref{Dnfinal}) with the numerically evaluated
splittings for the first three pairs of levels. Within our numerical
precision, we always find the zeros of $\Dta_n$ to agree with
Eq.~(\ref{quench2}). Note, however, that for other values of $H_c$,
the discrepancy between the numerics and Eq.~(\ref{Dnfinal}) is well
outside our numerical error, so that Eq.~(\ref{Dnfinal}) is not exact,
even though as an asymtotic estimate of the splitting it is rather
good. This means that the leading semiclassical approximation does not
give the eigenvalues themselves exactly, and only the quenching points
appear to be so reproduced. The second point to note is that for
$n=1$ (the pair of first excited states in each well), the highest
field quenching point is lost, for $n=2$, the highest two points are
lost, and so on, exactly as indicated by Eq.~(\ref{quench2}).

Next, let us compare our answers with previous work. Let us consider
the Gamow factor $\Gam_0$ first. Except for the replacement of $J$
by $\baj$ and $h_{x0}$ by $h_x$, this is precisely the action in
Eq.~(3.10) of \rno{opus}(c). This agreement is unsurprising,
because if we write $\Dta_n$ in
the form of a prefactor $c_1$ times a Gamow factor\
$\exp(-J c_2)$ where $c_2 = O(1)$, then the $J\to\baJ$, and
$h_{x0} \to h_x$ corrections in Eq.~(\ref{Sr}) represent terms that
should be included in the prefactor $c_1$, which we did not seek
to find in \rno{opus}(c). The detailed form of the prefactor
is perhaps more interesting. Up to multiplicative terms of order
$J^0$, our answer for $\Dta_n$ agrees precisely with
that in \rno{klzp,lkpp}. We do not understand, however, how
these papers have succeeded in sidestepping the difficulties in the
path integral treatment that were noted by Enz and Schilling \cite{es},
and by Belinicher, Providencia,
and da Providencia \cite{bpp}. In \rno{klzp}, for instance,
the problem is treated by writing the spin coherent state expectation
value of the Hamiltonian (\ref{ham}) in spherical polar coordinates,
and integrating out $\cos\tta$ (the $J_z$ projection) exactly, and
then addressing the resulting effective Lagrangian for the $\phi$
coordinate exactly as for a massive particle in one dimension.
In performing the integration over $\tta$, however, it is not clear
to us why $S^2$ is replaced by $S(S+1)$ in the scalar potential
$V(\phi)$ [see Eq.~(12) there], but not in the vector potential
$\Tta(\phi)$.

A related point, which is relatively minor, but has scope for
creating confusion, is that if the Gamow factor is written as
$\exp(-J c_2)$ with $c_2 = O(1)$, then it is safest to write
the $J$ dependence of the prefactor as $\om_0 J^{n+1/2}$, since
$\om_0$ depends on parameters such as $k_1$ and $k_2$, whose scaling
with $J$ is a matter of choice, at least as far as model Hamiltonians
are concerned.

One further check is obtained by considering the limiting case
$H_x = 0$, answers for which are known. [See, e.g., Eq.~(16) of
\cite{es}, Eq.~(48) of \cite{bpp}, or \cite{jmp1}.]
Transcribing Eqs.~(4.30) and (4.31) from \cite{jmp1}
in terms of the present parameters, we get
\bea
\Dta_n &=& {1\over n!} F_0^n \Dta_0; \label{Dtahx0} \\
F_0 &=& 8J{\sqrt\lam \over 1 - \lam}, \label{F0} \\
\Dta_0 &=& 8\om_{00} \lp{J \over \pi}\rp^{1/2}
           {\lam^{1/4} \over 1 + \sqrt\lam}
        \lp{ 1 - \sqrt\lam \over 1 + \sqrt\lam}\rp^J, \label{Dta00}
\eea
with $\om_{00} = 2J(k_1 k_2)^{1/2}$. It follows from
Eqs.~(\ref{om0fe8}--\ref{Lamtot}) that as $H_x \to 0$,
$\om_0 \to \om_{00}$, $F \to F_0$ [see
Eq.~(\ref{Ffe8})], $\cos\Lam_n \to \pm 1$, and
\beq
\Gam_0 \to \lp J + \tshf \rp
            \ln {1 +\sqrt\lam \over 1 - \sqrt\lam}.
         \label{limGam}
\eeq
It is then easy to see that our present answers for $\Dta_n$
go over precisely into Eqs.~(\ref{Dtahx0})--(\ref{Dta00}).

\subsection{Evaluation of Action Integrals}
\label{details}
To carry out the evaluation of Eqs.~(\ref{Phipr})--(\ref{Q1def}),  
it is convenient to measure energies (including $\om_0$) in units of
$k_1\baj^2$, and introduce the scaled variable $\mu = m/\baj$.
In terms of these variables, we have
\bea
w(m) &=& (1+\lam)(1- \mu^2)/2, \label{wofm} \\
t_1(m) &=& -h_x (1-\mu^2)^{1/2}, \label{t1ofm} \\
t_2(m) &=& (1-\lam)(1 - \mu^2)/4. \label{t2ofm}
\eea

The turning points $\mu_0 = m_0/\baj$, and $\mu_1 = m_1/\baj$
(for $\eps = 0$) are given by
\bea
\mu_0 &=& (1-h_x^2)^{1/2}, \label{mu0} \\
\mu_1 &=& [(1-\lam - h_x^2)/(1-\lam)]^{1/2}. \label{mu1}
\eea
It is most convenient to express everything in terms of $\mu_0$ and
$\mu_1$, so we give inverse formulas as well:
\bea
h_x = (1- \mo0)^{1/2}, \label{hx} \\
\lam = (\mo0 - \mo1)/(1 - \mo1). \label{lam}
\eea
The mass and the small oscillation frequency are given by
\bea
M = {1 \over 2\lam h_x^2}
        = \hf {1-\mo1 \over (1-\mo0)(\mo0 - \mo1)},
                        \label{M1} \\
\om_0 = 2 [\lam(1-h_x^2)]^{1/2}/\baj
         = {2\mu_0 \over \baj}
              \lp{\mo0 - \mo1 \over 1 - \mo1}\rp^{1/2}.
                        \label{om0}
\eea

To evaluate the integrals, we need expressions for $\kap_0$,
$\chi_0$, $\kap'_0$, etc. in the ranges $\mu_1 < \mu < \mu_0$,
and $0 < \mu < \mu_1$. The requisite calculations are straightforward
so we give the main results only. First, in the range
$\mu_1 < \mu < \mu_0$, we get
\bea
B_0 &=& \cosh \kap_0 = 
          {1 - \mo1 - [(\mo0 - \mo1)(\mu^2 - \mo1)]^{1/2}
              \over [(1 - \mo0) (1 - \mu^2)]^{1/2}}, \label{kgb1} \\
\sqrt{B^2_0 - 1} &=& \sinh \kap_0 = 
          { \lp\sqrt{\mo0 - \mo1} - \sqrt{\mu^2 - \mo1}\rp
              \sqrt{1 - \mo1} 
              \over [(1 - \mo0) (1 - \mu^2)]^{1/2}}, \label{kgb2} \\
B'_0 &=& - \hf { 1 - \mo1 \over
             [(1-\mo0)(1-\mu^2)(\mo0 - \mo1)
                       (\mu^2 - \mo1)]^{1/2}}, \label{kgb3} \\
{\om_0 B'_0 \over \sqrt{B_0^2 - 1}} &=&
         -{\mu_0 \over \baj}
             {1   \over \sqrt{\mu^2 - \mo1}
            \lp\sqrt{\mo0 - \mo1} - \sqrt{\mu^2 - \mo1}\rp}. \label{iQ11}
\eea
Next, in the range, $0 < \mu < \mu_1$, we first put $\eps=0$ in
Eqs.~(\ref{chcos}) and (\ref{shsin}), and solve, to obtain
\bea
\cosh\kap_0 = [(1-\mo1)/(1-\mo0)]^{1/2}, \label{kgb5} \\
\cos\chi_0 = [(1-\mo1)/(1-\mu^2)]^{1/2}. \label{cc0} \\
\sin\chi_0 = [(\mo1 - \mu^2)/(1 - \mu^2)]^{1/2} \label{sc0}
\eea
We then differentiate Eqs.(\ref{chcos}) and (\ref{shsin}), and
set $\eps = 0$ to obtain the equations
\bea
\sinh\kap_0 \cos\chi_0\ \kap'_0 
          - \cosh\kap_0 \sin\chi_0\  \chi'_0 &=& 0, \label{der1} \\
\cosh\kap_0 \sin\chi_0\  \kap'_0 
          + \sinh\kap_0 \cos\chi_0\  \chi'_0 &=&
                -1/8t_2\sinh\kap_0 \sin\chi_0.
                                  \label{der2}
\eea
Solving these, we obtain
\beq
\left( \begin{array}{c} \kap'_0 \\ \chi'_0 \end{array} \right)
    = -{(1-\mo1)^{1/2} \over 2(\mo0 - \mu^2)}
       \lp \begin{array}{c}
          (\mo0 - \mo1)^{-1/2} \\ (\mo1 - \mu^2)^{-1/2}
           \end{array} \rp. \label{derans}
\eeq

The first integral that we wish to evaluate is $\Gam_0$, which will
give us the dominant WKB or Gamow factor in the tunnel splitting.
We divide the integral into two parts by breaking the integration
range at $m_1$. From the right-hand part, an integration by parts
gives
\bea
\Gam_{01} &=& 2\baj \intrt \kap_0 d\mu \nnu \\
    &=& 2\baj \left[ \kap_0(\mu)\mu \!\bigm|_{\mu_1}^{\mu_0}
                - \intrt {\mu \over \sinh\kap_0}
                        {dB_0(\mu) \over d\mu} d\mu \right],
       \label{Gam01}
\eea
while from the left-hand part we get
\beq
\Gam_{02} = 2\baj \int_0^{\mu_1} \kap_0 d\mu 
            = 2 \baj  \kap_0(\mu_1)\mu_1, \label{Gam02}
\eeq
as $\kap_0$ is a constant in this range. Since
$\kap_0(\mu_0) = 0$, $\Gam_{02}$ cancels
the first term in Eq.~(\ref{Gam01}), leaving us only with the
second for $\Gam_0$. Using Eqs.~(\ref{kgb1}) and (\ref{kgb2}),
we find
\beq
\Gam_0 = 2\baj (1-\mo1)^{1/2} \intrt
                 {d\mu \over (1-\mu^2)(\mu^2 - \mo1)^{1/2}}.
         \label{Gam02nd}
\eeq
The integration is now elementary, and the result, expressed
back in terms of $\lam$ and $h_x$ is Eq.~(\ref{Sr}).

The second integral to be evaluated is $\Lam_n$. For the first term
in Eq.~(\ref{Lamans}), we integrate by parts, and use
Eqs.~(\ref{cc0}) and (\ref{sc0}):
\bea
2\baj\intcen \chi_0(\mu)d\mu &=&
         2\baj\intcen {\mu\over \sin\chi_0}{d\over d\mu}\
                \cos\chi_0\ d\mu \nnu\\
   &=& 2\baj \intcen {\mu^2 \over (1-\mu^2)(\mo1- \mu^2)^{1/2}}
                 d\mu \nnu \\
   &=& \pi\baj [1 - (1-\mo1)^{1/2}]. \label{Lam1}
\eea
For the second term in Eq.~(\ref{Lamans}), we have, with
$\eps = (n+\tshf)\om_0$, and Eqs.~(\ref{om0}) and (\ref{derans}),
\bea
2\eps\baj\intcen \kap'_0 d\mu &=&
        -(2n + 1) \mu_0 (\mo0 - \mo1)^{1/2}
         \intcen {d\mu \over (\mo0 - \mu^2)(\mo1 - \mu^2)^{1/2}}
                 \nnu \\
         &=& -\lp n + \tshf \rp \pi. \label{Lam2}
\eea
Adding together the parts, and rewriting the result in terms of
$H_x$ and $\lam$, we get Eq.~(\ref{Lamtot}). The restriction that
$\Lam_n$ be positive follows from the fact that we chose $q(m)$ to
have a positive real part in Eq.~(\ref{Dgen3}).
Thus $\Lam$ is necessarily
positive as defined in Eq.~(\ref{Lamans}). If $H_x$ is so large
as to yield a negative value for the function of $H_x$ that results
after doing the integral, that means that in fact there are
no irregular turning points in the problem. Both terms in
Eq.~(\ref{Dgen3}) are then equal, and the formula reduces to the
expected one when there are only regular turning points.

Note that unlike Eq.~(\ref{Sr}), what appears in Eq.~(\ref{Lamtot})
is the ratio $H_x/H_c$, i.e., $h_{x0}$, not $h_x$. This fact is
important for the location of the diabolical points.

The third integral we need is that of $\kap'_0$ from $0$ to $\mu_1$.
Using Eqs.~(\ref{om0}) and (\ref{derans}), we get
\beq
2\om_0\baj\intcen \kap'_0 d\mu
     = -2\mu_0 \intcen {d\mu \over \mo0 - \mu^2}
     = \ln {\mu_0 - \mu_1 \over \mu_0 + \mu_1}. \label{int3}
\eeq

The fourth and last integral needed is $Q_1$. Substituting
Eq.~(\ref{iQ11}) in Eq.~(\ref{Q1def}), we obtain
\beq
Q_1 = -\int_{\mu_0}^{\mu_1}
               \left[ {\mu_0   \over \sqrt{\mu^2 - \mo1}
            \lp\sqrt{\mo0 - \mo1} - \sqrt{\mu^2 - \mo1}\rp}
            - {1 \over \mu_0 - \mu} \right] d\mu. \label{Q1fe8}
\eeq
The integrand is now nonsingular at $\mu = \mu_0$. We can make this
manifest by rationalizing the difference of square roots in the first
term. Some simple algebra yields
\beq
Q_1 = - \int_{\mu_0}^{\mu_1} {1 \over \sqrt{\mu^2 - \mo1}}
          {\mo0 + \mu^2 - \mo1 \over
           \mu_0\sqrt{\mo0-\mo1} + \mu\sqrt{\mu^2 - \mo1}}
         d\mu. \label{Q12}
\eeq
We now make the substitution $\mu = \mu_1\cosh z$, and define
\beq
\cosh z_0 = \mu_0/\mu_1 .\label{z0def}
\eeq
This yields
\bea 
Q_1 &=& -\int_0^{z_0}
           {\cosh 2z_0 + \cosh 2z \over
            \sinh 2z_0 + \sinh 2z} dz \nnu \\
    &=& -\int_0^{z_0} {\cosh(z+z_0) \over \sinh(z+z_0)} dz
          = -\ln(2\cosh z_0) \nnu \\
    &=& -\ln{2\mu_0 \over \mu_1}. \label{Q1ans}
\eea

We now have all the ingredients needed to calculate the quantity
$F$. Substituting Eqs.~(\ref{M1}), (\ref{om0}), (\ref{int3}),
and (\ref{Q1ans}) in Eq.~(\ref{Fans}), and writing the result
in terms of $\lam$ and $h_x$, we obtain Eq.~(\ref{Ffe8}). Note that
in writing down the final answer, we have replaced $\baj$ by $J$
and $h_x$ by $h_{x0}$ in this formula.
This is because $F$ is part of the pre-exponential factor in
$\Dta_n$, which is determined only to leading order in $1/J$.
Keeping higher order corrections by distinguishing between
$\baj$ and $J$ or $h_x$ and $h_{x0}$ is not justified.

The final answer (\ref{Dnfinal}) for $\Dta_n$ is obtained by
substituting
Eqs.~(\ref{Sr}), (\ref{Lamtot}), and (\ref{Ffe8}) in
Eq.~(\ref{Dgen4}).

\acknowledgments
This work is supported by the NSF via grant number DMR-9616749.
I am indebted to Wolfgang Wernsdorfer and Jacques Villain for
useful discussions and correspondence about \Fe8.

\begin{figure}
\caption{Spectrum of the Hamiltonian (\ref{ham}) for $J=3$, as a function of
$H_x/H_c$. $H_z/H_c = 0$, $0.07454$, and $0.1491$ in (a), (b), and (c)
respectively.  The small ovals indicate points that are narrowly avoided
anticrossings, but appear to be crossings on low resolution.}
\label{fig1}
\end{figure}

\begin{figure}
\caption{Critical energy curves for the Hamiltonian
(\ref{ham}), showing the dual labelling scheme.}
\label{fig2}
\end{figure}

\begin{figure}
\caption{Magnified view of the lower left hand region
of \fno2 showing the point of tangency $m^*$ between
$U_0$ and $U_*$, and turning points at $m=-m_t$ and
$-m_1$ for an energy $E$.}
\label{fig3}
\end{figure}

\begin{figure}
\caption{Comparison between numerical (solid lines) and
analytic [Eqs.~(\ref{Dnfinal}--\ref{Lamtot}), dashed lines]
results for the splitting between the first three pairs of
levels for $H_z = 0$. The parameters are $k_1 = 0.321\ $K,
$k_2=0.229\ $K, close to those for \Fe8.}
\label{fig4}
\end{figure}


\begin{references}
\bibitem{fn1} Since the eikonal equation (\ref{hjeq}) has four
solutions, there are four DPI solutions, all of which break down
at the turning point at $m=-m_1$. In \rno{jmp2} we derive
connection formulas for all four solutions, of which,
Eq.~(\ref{Crite}) is the case there designated by $\sig_1 = -1$,
$\sig_2 = 1$.
\bibitem{fn2} The result in \rno{ll} is missing the correction
factor $g_n$, as mentioned in footnote \cite{bm}.  For the ground
state this factor is well known and understood, and an elementary
discussion is given in \rno{amjp}.
\end{references}
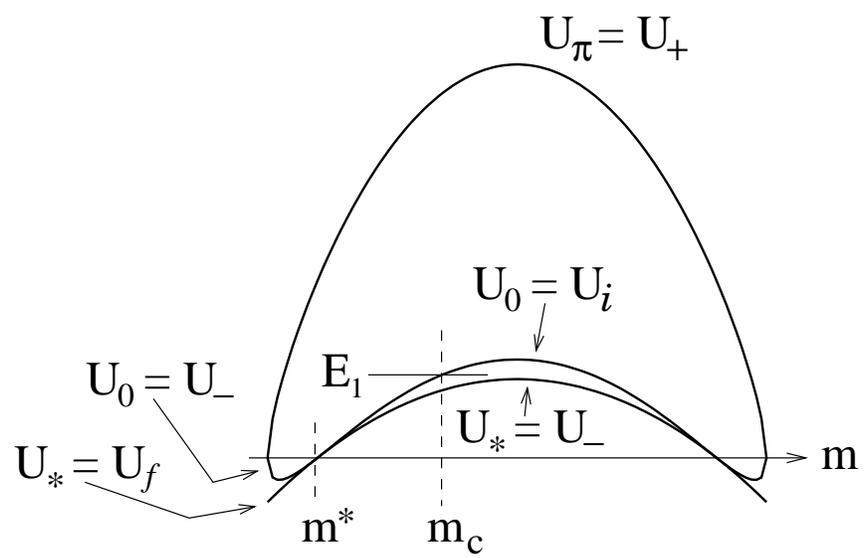
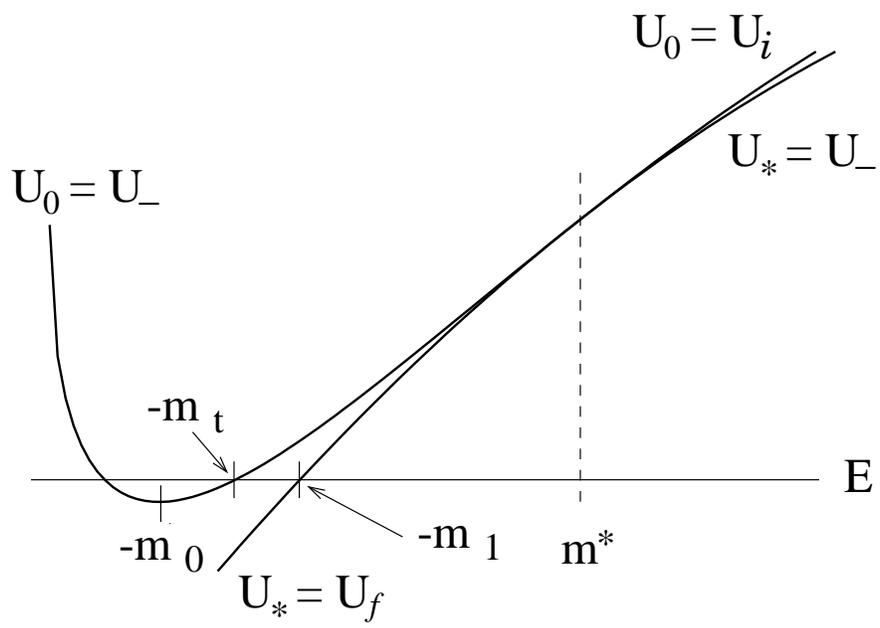
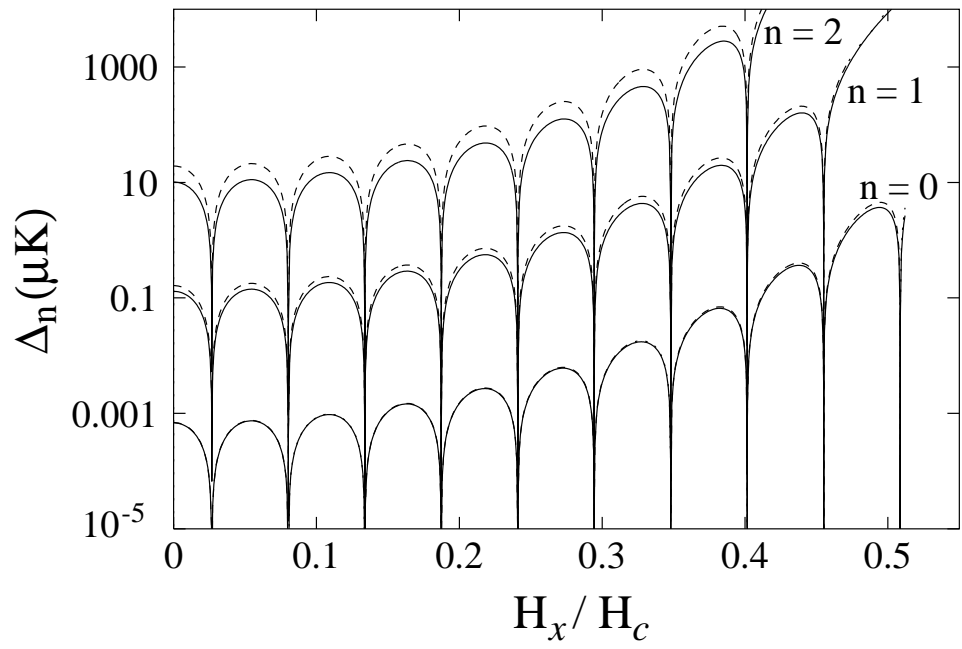
\end{document}